\documentclass[a4paper]{article}

\usepackage{INTERSPEECH2021}
\usepackage{multirow}

\title{Improved far-field speech recognition using Joint Variational Autoencoder}
\name{Shashi Kumar$^1$, Shakti P. Rath$^2$, and Abhishek Pandey$^1$ }
\address{
  $^1$Samsung R\&D Institute India - Bangalore\\
  $^2$Reverie Language Technologies, India}
\email{sk.kumar@samsung.com, shakti.rath@reverieinc.com,abhi3.pandey@samsung.com}

\begin{document}

\maketitle
\begin{abstract}
Automatic Speech Recognition (ASR) systems suffer considerably when source speech is corrupted with noise or room impulse responses (RIR).
Typically, speech enhancement is applied in both mismatched and matched scenario training and testing.
In matched setting, acoustic model (AM) is trained on dereverberated far-field features while in mismatched setting, AM is fixed.
In recent past, mapping speech features from far-field to close-talk using denoising autoencoder (DA) has been explored.
In this paper, we focus on matched scenario training and show that the proposed joint VAE based mapping achieves a significant improvement over DA.
Specifically, we observe an absolute improvement of $2.5\%$ in word error rate (WER) compared to DA based enhancement and $3.96\%$ compared to AM trained directly on far-field filterbank features.
\end{abstract}

\section{Introduction}
The performance of automatic speech recognition (ASR) systems have improved greatly in close-talking scenario but it suffers heavily when tested in far-field condition.
Far field speech recognition is a challenging problem because of various convolutive and additive distortions like room impulse responses (RIR), background noise etc.
One of the dominant approaches to handle this problem is speech enhancement in which the received speech is dereverberated to reduce such distortions.
In neural network based single-channel speech enhancement techniques, the enhancement model is trained either in matched or mismatched scenario. In matched scenario, the enhancement model is trained jointly with acoustic model (AM) whereas in mismatched scenario, at first an AM is trained on close-talk speech and then speech enhancement model is trained separately to reduce dereverberation which is decoded using the AM trained earlier. In this paper, we focus on matched scenario training of single-channel speech enhancement models.

Speech enhancement has been an important and widely researched problem in far-field speech recognition. 
Several approaches are proposed in the direction of domain adaption to enhance far-field speech \cite{Glass-AMI-review,orig-MSE-speechEnhance,phase-based-mask,phase-fact,DA4-properDNNbasedCleaning-MSEloss,domain-adaptation-1, domain-adaptation-2, DA1-noiseAware}.
Other methods for speech enhancement include spectral magnitude based approaches that estimate an inverse-filter to cancel the effect the late reverberation \cite{WPE-orig-paper,WPE-DNN-2017} or non-negative matrix factorization \cite{NMF-based}.  
Moreover, different mask based approaches have also been employed for dereverberation and noise-reduction \cite{cIRM-icassp-2017,phase-based-mask}.
The most dominant work in speech enhancement include mapping characteristics of speech from the source domain to target domain \cite{DA1-noiseAware,DA2-maskBased,DA4-properDNNbasedCleaning-MSEloss,DA5-rnnbased-MSE,DNN-based-enhance-multiCondition}, commonly known as denoising autoencoder (DA). 
In \cite{microsoft-matched-2016}, multiple architectures for matched scenario training with different intermediate outputs have been explored.
DA has also been explored for speaker recognition in far-field scenario \cite{DA-spkRecog}. 
In another work, bottleneck feature mapping is explored \cite{DA6-bottleNeck-MSE} which maps bottleneck features from the source domain to that of the target domain. 
One of the common approaches for speech enhancement is to map the speech features from far-field domain to close-talk domain using a neural network known as denoising autoencoder (DA).
In a typical far-field setting, the mapping from reverberant to clean speech can be highly non-linear and time-variant that can not be optimally represented by a stationary DA.

Variational Autoencoder (VAE) \cite{vae-orig-paper} is a class of generative model that projects input space to a latent space using an encoder and then reconstructs the original input using a decoder. 
VAEs have been explored for many tasks like speech transformation from a source domain to target domain \cite{glass-vae-data-aug}, showing orthogonality of speech attributes that can help in domain translation \cite{glass-vae-ortho}, speaker verification \cite{glass-vae-factorized-spk}.
In \cite{dataaug_glass} it has been applied for data augmentation where a speech transformation is learned to transform the data from the source domain to the target domain without altering the linguistic content to create additional transcribed data. The pooled data is used for acoustic model training that showed substantial improvement in ASR accuracy.
In \cite{glass-vae-ortho} it is shown that different speech attributes, such as speaker characteristics and linguistic content, are mutually orthogonal in the latent space. 
Leveraging such orthogonality, VAE is applied for speech transformation, where speaker characteristics are changed without changing linguistic content  and vice versa.
Later VAE was extended to a factorized hierarchical representation and applications were shown for speaker verification and speech recognition \cite{fhvae,asr_glass}.
VAE has also been explored for voice activity detection \cite{vae-vad}.
Many existing speech enhancement techniques leverage availability of parallel data (aligned far-field and close-talk features) for dereverberation with the help of deep neural networks.
The conventional VAE is a completely unsupervised technique that is designed for probabilistic generative modeling without taking advantage of parallel data.
Also, mapping far-field features to close-talk features can not be done by a conventional VAE because of constraint that input and output must be same mathematically.

Recently, we proposed a novel method for speech enhancement, termed as joint VAE \cite{joint-VAE-paper}, which showed promising results in mapping distant to close-talk speech in mismatched scenario.
The constraint with conventional VAE is addressed in joint VAE by learning a joint-distribution of far-field and close-talk features for a common latent space and the resulting variational lower bound (ELBO) is maximized to train the inference and generative network parameters.
Another difference between VAE and joint-VAE is that the former optimizes an ELBO consisting of two terms, namely, a reconstruction error and KL-divergence, whereas joint-VAE involves two reconstruction errors and KL-divergence.
The model involves a common encoder network for inference that takes far-field features as input and two decoder networks that reconstructs the close-talk and far-field features separately.
In this paper, we propose to explore joint VAE as an alternative to DA for speech enhancement in matched scenario and show that it yields consistent and considerable improvement in ASR accuracy compared to DA and AM trained on far-field speech.
We also propose to relax approximation made in modeling posterior distribution in joint VAE \cite{joint-VAE-paper} and propose a suitable extension of joint VAE architecture for the same.
This relexation of approximation gives a significant improvement.

The remainder of the paper is organized as follows. In Section~\ref{sec:review}, we review  conventional VAE, joint VAE and deduce the final loss function corresponding to matched scenario training. In Section~\ref{sec:exp-res}, the experimental results on the AMI dataset are presented. Conclusion and future work are presented in Section~\ref{sec:conclusions}.


\section{Review of Variational Autoencoder and Joint VAE}
\label{sec:review}
I
The VAEs are essentially an encoder-decoder based model where encoder maps input feature space to latent space and decoder tries to reconstruct the features given samples from latent space.
Standard VAE tries to reconstruct the input space and hence it does not offer domain translation. 
Recently, we proposed joint VAE \cite{joint-VAE-paper} which enables domain transformation by learning a joint distribution of two domains for a common latent space.
Details of standard VAE and joint VAE are explained in following sections.

\subsection{Variational Autoencoder (VAE)}
\label{sec:conventional-VAE}
The underlying principle in VAE is to assume that the observed data has been generated by a random process that involves latent variables.
Let the sequence of latent variables be denoted by $\pmb{z}_1, \pmb{z}_2, \cdots, \pmb{z}_N$ and the observed data be denoted by $\pmb{x}_1, \pmb{x}_2, \cdots, \pmb{x}_N$.
In order to model the observed process, it is necessary to estimate the the posterior distribution $p_{\theta}(\mathbf{z}|\mathbf{x})$ given samples from input space distribution,
where $p_{\theta}$ denotes family of distributions parameterized by $\theta$. Using Baye's rule $p_{\theta}(\mathbf{z}|\mathbf{x)} = p_{\theta}(\mathbf{x}|\mathbf{z})p_{\theta}(\mathbf{z})/p_{\theta}(\mathbf{x})$.
In practice $p_{\theta}(\mathbf{z}|\pmb{x})$ becomes intractable even for a simpler distribution family.
So it is approximated by another parametrized distribution by minimizing Kullback-Leibler (KL) divergence between the two distributions. 
Let the other distribution be denoted by $q_{\phi}(\pmb{z}|\pmb{x})$.
It is straight forward to show that following relation holds
\begin{align}
 \log\:p_{\theta}(\pmb{x}) &= \mathcal{L}_1(\theta,\phi;\pmb{x}) + KL(q_{\phi}(\pmb{z}|\pmb{x})\:||\:p_{\theta}(\pmb{z}|\pmb{x})) \\
 &\ge  \mathcal{L}_1(\theta,\phi;\pmb{x})
\end{align}
where 
$p_{\theta}(\pmb{x})$ denotes the marginal distribution of the observed data and 
$\mathcal{L}_1(\theta,\phi;\pmb{x})$ is called the variational lower bound, which is defined as
\begin{align}
\label{eq:lowerbound-vae}
\mathcal{L}_1(\theta,\phi;\pmb{x})  &= \int_{z} q_{\phi}(\pmb{z}|\pmb{x})\:\log\frac{p_{\theta}(\pmb{x},\pmb{z})}{q_{\phi}(\pmb{z}|\pmb{x})}\\
&= E_{q_{\phi}(\pmb{z}|\pmb{x})} \log p_{\theta}(\pmb{x}|\pmb{z}) - KL \left( q_{\phi}(\pmb{z}|\pmb{x}) || p_{\theta}(\pmb{z}) \right) \nonumber
\end{align}
Commonly, the prior $p_{\theta}(\pmb{z})$ is modeled by isotropic Gaussian distribution $p_{\theta}(\pmb{z}) = \mathcal{N}(\pmb{z}; \mathbf{0}, \mathbf{I})$
and the distributions $q_{\phi}(\pmb{z}|\pmb{x})$ and $p_{\theta}(\pmb{x}|\pmb{z})$ by diagonal Gaussian distributions which are represented by neural networks.
Parameters $\phi$ and $\theta$ of the these distributions are jointly estimated by minimizing negative of the variational lower bound (Eq~\ref{eq:lowerbound-vae}).
To compute expectation term in Eq~\ref{eq:lowerbound-vae} of variational lower bound, samples $\hat{\pmb{z}}$ needs to be drawn from the posterior $q_{\phi}(\pmb{z}|\pmb{x})$. 
Since sampling is a  non-differentiable operation, the standard error backpropagation cannot directly be applied for the training. 
To handle this limitation, the re-parameterization trick \cite{vae-orig-paper} is used to make the sampling operator differentiable.

It is important to note here that it may appear that conventional VAE may be extended for domain conversion.
In past, it has been explored for speech enhancement task (denoising VAE (DVAE)) where conventional VAE is applied to learn a mapping from noisy to clean speech domains.
However from Eq.~\ref{eq:lowerbound-vae}, it may be noted that in VAE, the input and output random processes must be the same, i.e., $\pmb{x}$.
If the input and output are forced to be different, as in the case of DVAE, the results may become unpredictable.
Therefore from a theoretical point of view, such domain conversion cannot be justified within the premises of conventional VAE.
For this reason, results are not shown for DVAE in this paper.

\subsection{Joint Variational Autoencoder (Joint VAE)}
\label{sec:joint-VAE}
In this section, we present a detailed description of joint VAE \cite{joint-VAE-paper}. 
For consistency, we denote far-field features or input domain by $\pmb{x}$ and output domain or close-talk features by $\pmb{y}$.
The motivation is to learn mapping from $\pmb{x}$ to $\pmb{y}$ in a time synchronous fashion.
We assume that we have access to parallel data of these domains aligned in time at training phase. 
In joint VAE, the distribution of the data from input domain and output domains are modeled using a joint probability distribution, and the variational lower bound is re-defined as follows
\begin{align}
\mathcal{L}_{2}(\theta,\phi;\pmb{x},\pmb{y}) &= \int_{\pmb{z}} q_{\phi}(\pmb{z}|\pmb{x},\pmb{y})\log\frac{p_{\theta}(\pmb{x},\pmb{y},\pmb{z})}{q_{\phi}(\pmb{z}|\pmb{x},\pmb{y})} \nonumber\\
                &= E_{q_{\phi}(\pmb{z}|\pmb{x})}\log\:p_{\theta}(\pmb{x}|\pmb{z}) + E_{q_{\phi}(\pmb{z}|\pmb{x})}\:\log\:p_{\theta}(\pmb{y}|\pmb{x},\pmb{z}) \nonumber\\
                & \hspace{2cm} - KL(q_{\phi}(\pmb{z}|\pmb{x})\:||\:p_{\theta}(\pmb{z}))
\label{eq:joint_vae_lb}
\end{align}
The modified lower bound consists of two conditional distributions $p_{\theta}(\pmb{y}|\pmb{x},\pmb{z})$ and $p_{\theta}(\pmb{x}|\pmb{z})$ and the posterior distribution $q_{\phi}(\pmb{z}|\pmb{x})$, each of which is represented using a neural network. 
in \cite{joint-VAE-paper}, an approximation is made $q_{\phi}(\pmb{z}|\pmb{x},\pmb{y}) = q_{\phi}(\pmb{z}|\pmb{x})$ assuming the mapping between domains $\pmb{x}$ and $\pmb{y}$ is deterministic.
All the above conditional distributions are modeled by a diagonal Gaussian distribution 
and the prior $p_{\theta}(\pmb{z})$ is modeled by isotropic Gaussian. The neural network parameters are jointly optimized by minimizing negative of modified lower bound (Eq.~\ref{eq:joint_vae_lb}).
In practice, the actual loss that is used to train networks is given by
\begin{align}
\label{eq:loss-proposed}
\mathcal{L}_3 =  \lambda_{1}\:\text{MSE}_\text{x} + \lambda_{2}\: \text{MSE}_\text{y} + \lambda_{3}\: \text{KLD},
\end{align}
where the first term is $\text{MSE}_\text{x}$ is heteroscedastic MSE \cite{hetero-shashi} between input $\pmb{x}$ and reconstructed $\pmb{x}$ output, the second term $\text{MSE}_\text{y}$ is heteroscedastic MSE between true $\pmb{y}$ and reconstructed $\pmb{y}$ output.
The third term is KL-divergence between $q_{\phi}(\pmb{z}|\pmb{x})$ and prior distribution $p_{\theta}(\pmb{z})$.
In the joint VAE loss, the role of the the KLD term is to smoothen the decision boundaries among different classes. It forces the distribution $q_{\phi}(\pmb{z}|\pmb{x})$ to be as close to isotropic diagonal Gaussian and it induces inherent disentanglement \cite{beta-vae}, whereas the reconstruction terms encourage deviation from prior distribution in the latent space so as to encode data effectively in different dimension of latent variables.

In contrast to conventional VAE, joint VAE consists of one encoder and two decoders.
The last LSTM layer of the encoder is followed by two parallel fully connected layers with linear activation, predicting mean and log-variance.
Similarly, the lower decoder network consists of two parallel fully connected layers with linear activation, predicting mean and log-variance. 
It takes $\hat{\pmb{z}}$ as input and predicts mean and log-variance of $\pmb{x}$.
The upper decoder takes $\hat{\pmb{z}}$ and $\pmb{x}$ as input and predicts $\pmb{y}$.

We now propose to relax the approximation $q_{\phi}(\pmb{z}|\pmb{x},\pmb{y}) = q_{\phi}(\pmb{z}|\pmb{x})$.
Now, the posterior distribution, modeled by a neural network, requires both $\pmb{x}$ and $\pmb{y}$ as input.
Unfortunately, time-aligned IHM features, $\pmb{y}$, are not available at test time.
We propose to use DA based mapping network which maps SDM features to IHM features at input side before encoder.
Thus, SDM features with predicted IHM features using DA is given as input to encoder network.
This DA network is trained jointly with joint VAE model by minimizing standard mean square error (MSE) between true and predicted IHM features. Now, the final loss is given by
\begin{align}
\label{eq:loss-proposed-woApprox}
\mathcal{L}_3 =  \lambda_{1}\:\text{MSE}_\text{x} + \lambda_{2}\: \text{MSE}_\text{y} + \lambda_{3}\: \text{KLD} + \lambda_{DA}\:\text{MSE}_{DA}
\end{align}
where $\text{MSE}_\text{DA}$ is standard MSE to train the DA network.

\subsection{Joint training with AM}
\label{sec:jointTrain}
In matched scenario, we propose to train joint VAE model jointly with AM.
The mean of predicted IHM features $\pmb{y}$, denoted by $\mu_{IHM}$, is given as input to AM and trained jointly by minimizing cross entropy (CE) loss. Now, the final loss is given by
\begin{align}
\label{eq:loss-proposed-woApprox-joint}
\mathcal{L}_4 = \mathcal{L}_3 + \beta\:\text{CE}
\end{align}
where $\mathcal{L}_3$ is final loss which is used to train joint VAE model given by Eq \ref{eq:loss-proposed}, when the posterior distribution is approximated by excluding $\pmb{y}$ from input, and Eq \ref{eq:loss-proposed-woApprox} when the approximation is relaxed. 
%


\section{Experiments and Results}
\label{sec:exp-res}

\subsection{Experimental Setup}
\label{sec:baseline}
We conducted experiments on AMI dataset \cite{ami} where far-field and close-talk speech parallel data is available.
It consists of around 100 hours of meeting speech recordings of non-native speakers. 
The recordings are in English using both individual head microphones (IHM) and one or more distant microphones.
In our experiments, we use audios from IHM (close-talk) and the first distant microphone, referred as single distant microphone (SDM, far-field) which are time-aligned using beam forming.
We report the word error rate (\%WER) on standard dev set which is created by following Kaldi standard recipe for AMI corpus, it labels around 80 hours of data as training corpus and around 8 hours of data as standard dev set.
For the extent of this work, we use Kaldi \cite{kaldi} for GMM-HMM training and pytorch for joint VAE training.
We followed standard Kaldi recipe for AMI corpus to train LDA-MLLT-SAT GMM-HMM baseline system using IHM data.
We use this model to generate senone alignments for acoustic model training as it is well known that AM trained using senone alignments from IHM baseline outperforms the AM trained using senone alignments from SDM baseline \cite{microsoft-matched-2016}.

The ASR acoustic model is a LSTM-HMM system, which is trained using 41-dimensional log mel-filter bank features with $\pm 2$ splicing. The model consists of three LSTM layers with $512$ cells each and trained by minimizing cross entropy loss. We first train AM on IHM data which achieves $29.4\%$ WER. When this model is tested on SDM test set, the WER degrades to $70.03\%$. We then train AM on SDM data, considered as first baseline which achieves $55.52\%$ WER. Results are shown in Table~\ref{table:1}.
\begin{table}[h]
	\caption{WER (\%) on IHM and SDM dev set}
	\label{table:1}
	\centering
	\begin{tabular}{c c c}
		\toprule
		\textbf{Train} & \textbf{Target} & \textbf{WER}(\%) \\
		\toprule
		IHM & IHM & 29.3\\
		IHM & SDM & 70.03\\
		SDM & SDM & 55.52\\
		\bottomrule
	\end{tabular}
\end{table}

\subsection{Denoising Autoencoder (DA)}
\label{sec:da}
Our second baseline is a DA based speech enhancement model trained jointly with AM.
The DA maps SDM features to IHM features which are further spliced on-the-fly and passed to AM model to predict posterior probabilities.
The DA and AM are trained jointly by minimizing mean square error (MSE) between true and predicted IHM features and cross entropy loss.
Specifically, $\mathcal{L}oss =  \lambda_{1}\:\text{MSE} + \lambda_{2}\:\text{CE}$.
We use gridsearch to find best set of hyperparameters $\lambda_{1}$ and $\lambda_{2}$ for which values are picked from set $\{10^{-1}, 10^{0}, 10^{1}\}$.
We achieved $54.06\%$ as best WER. Comparison results are shown in Table~\ref{table:2}.

\subsection{joint VAE}
\label{sec:jvae}
In joint VAE architecture \cite{joint-VAE-paper}, the encoder network comprises of 3 LSTM \cite{lstm} layers.
The lower decoder subnetwork, $\text{Decoder}_\text{x}$, comprises of 2 LSTM layers.
The upper decoder subnetwork, $\text{Decoder}_\text{y}$, has same layer structure as $\text{Decoder}_\text{x}$.
The model is trained by minimizing the loss function defined in Eq~\ref{eq:loss-proposed}.
The value of hyperparameters $\lambda_{1}$, $\lambda_{2}$ and $\lambda_{3}$ are taken as $1$, $10$ and $0.1$ respectively, as mentioned in \cite{joint-VAE-paper}.
The acoustic model is trained jointly with joint VAE model using mean predicted by $\text{Decoder}_\text{y}$, implying that the ASR model is trained on predicted IHM features.
At test time, filterbank features are passed through joint VAE and the predicted IHM features are given to the acoustic model for decoding.

We first explore joint VAE model with approximation $q_{\phi}(\pmb{z}|\pmb{x},\pmb{y}) = q_{\phi}(\pmb{z}|\pmb{x})$.
As stated in \cite{joint-VAE-paper}, this approximated is made on assumption that mapping betwen far-field and close-talk features is deterministic and justified by reported results.
This variation of joint VAE model, is trained by minimizing the loss given by Eq \ref{eq:loss-proposed}, shows significant improvements in mismatched scenario \cite{joint-VAE-paper}.
In matched scenario, this variation of joint VAE is trained jointly with AM by minimizing loss given by Eq \ref{eq:loss-proposed-woApprox-joint} where $\mathcal{L}_3$ is given by Eq~\ref{eq:loss-proposed}. Unfortunately, it doesn't show any improvement in matched scenario, so we don't report results for the same.

We now relax the approximation and model $q_{\phi}(\pmb{z}|\pmb{x},\pmb{y})$ directly.
It can be observed that parallel IHM data is needed at both training and decoding time to learn the approximate posterior distribution $q_{\phi}(\pmb{z}|\pmb{x},\pmb{y})$.
Unfortunately, time-aligned IHM data is not available at decoding time.
To address this, we include a DA mapping network before encoder.
This DA maps features from SDM to IHM domain which is further spliced on-the-fly by $\pm 2$ frames and concatenated with input SDM features.
The concatenated feature is passed to encoder as input.
We use $2$ layer LSTM as DA.
Thus, the final model consists of DA in encoder side, joint VAE model and AM which is trained jointly by minimizing loss function described in Eq \ref{eq:loss-proposed-woApprox-joint} where $\mathcal{L}_3$ is given by Eq~\ref{eq:loss-proposed-woApprox}.

Results are shown in Table \ref{table:3}.
The final loss function, described in Eq \ref{eq:loss-proposed-woApprox-joint}, consists of $5$ hyperparameters.
We choose values of these hyperparameters from set $\{10^{-1}, 10^{0}, 10^{1}\}$.
To find the best set of hyperparameters, we choose the values depending on task based conjectures because grid search is infeasible.
Among empirical nuances, increasing values of $\lambda_{2}$ should improve predicted IHM features, increasing values of $\lambda_{DA}$ feeds better input to encoder which may lead to a better approximation of posterior $q_{\phi}(\pmb{z}|\pmb{x},\pmb{y})$.
Similarly, increasing $\beta$ should lead to a better AM.
Increasing weight of KLD ($\lambda_{3}$) leads to better factorization of latent space $\pmb{z}$ \cite{beta-vae} whereas decreasing the weight may lead to better reconstruction \cite{joint-VAE-paper}.
Since reconstruction of SDM features are not directly related with prediction of IHM features in the joint VAE model, we fix the value of $\lambda_{1}$ as $1$.
We first experiment with value of all hyperparmeters equal to $1$ which achieves $51.56\%$ as WER.
Following \cite{joint-VAE-paper}, we reduce the weight of KLD ($\lambda_{3}$) to $0.1$ but it did not improve the performance.
Keeping the same value of $\lambda_{3}$, we increase the value of $\lambda_{DA}$ to improve latent space but it further degrades the performance.
Now we increase the weight associated with IHM reconstruction error ($\lambda_{2}$) which could directly benefit the AM, it does improve the performance but the WER is still higher than the best model. This shows that increasing $\lambda_{2}$ improves the performance.
Learning from this, we now start another trail of experiments by setting $\lambda_{2}$ as $10$ and all other hyperparameters as $1$. This set of hyperparameters achieves $51.71\%$ as WER which is very close to the best model.
We further increase $\lambda_{DA}$ for better latent vectors but unfortunately it does not show a strong bearing on WER.
So, we revert $\lambda_{DA}$ back to $1$ and experiment with different values of $\lambda_{3}$. Decreasing the value of $\lambda_{3}$ degrades the performance slightly but increasing the value degrades the performance significantly.
Overall, the best WER we report is $51.56\%$ which is an absolute improvement of $2.5\%$ compared to DA based enhancement and $3.96\%$ compared to AM trained directly on SDM features.
\begin{table}[t]
	\caption{WER (\%) on SDM dev set using different enhancement models}
	\label{table:2}
	\centering
	\begin{tabular}{c c c}
		\toprule
		\textbf{Train} & \textbf{Model} & \textbf{WER}(\%) \\
		\toprule
		SDM & DA & 54.06\\
		SDM & joint-VAE & \textbf{51.56}\\
		\bottomrule
	\end{tabular}
\end{table}
\begin{table}[t]
\caption{WER (\%) of  joint-VAE on SDM dev set ($\lambda_{1}=1$)}
\label{table:3}
\centering
\begin{tabular}{c c c c c c}
\toprule
\textbf{Model} & $\mathbf{\lambda_{2}}$ & $\mathbf{\lambda_{3}}$ & $\mathbf{\lambda_{DA}}$ & $\mathbf{\beta}$ & \textbf{WER}(\%) \\[0.5ex]
\toprule
\multirow{9}{*}{Joint-VAE} & 1 & 1 & 1 & 1 & \textbf{51.56}\\
& 1 & 0.1 & 1 & 1 & 52.17\\
& 1 & 0.1 & 10 & 1 & 52.63\\
& 10 & 0.1 & 10 & 1 & 52.29\\
& 1 & 1 & 10 & 1 & 52.53\\
& 10 & 1 & 1 & 1 & 51.71\\
& 10 & 1 & 10 & 1 & 51.74\\
& 10 & 0.1 & 1 & 1 & 51.87\\
& 10 & 10 & 1 & 1 & 52.24\\
\bottomrule
\end{tabular}
\end{table}

\begin{figure}[t]
  \centering
  \includegraphics[width=\linewidth]{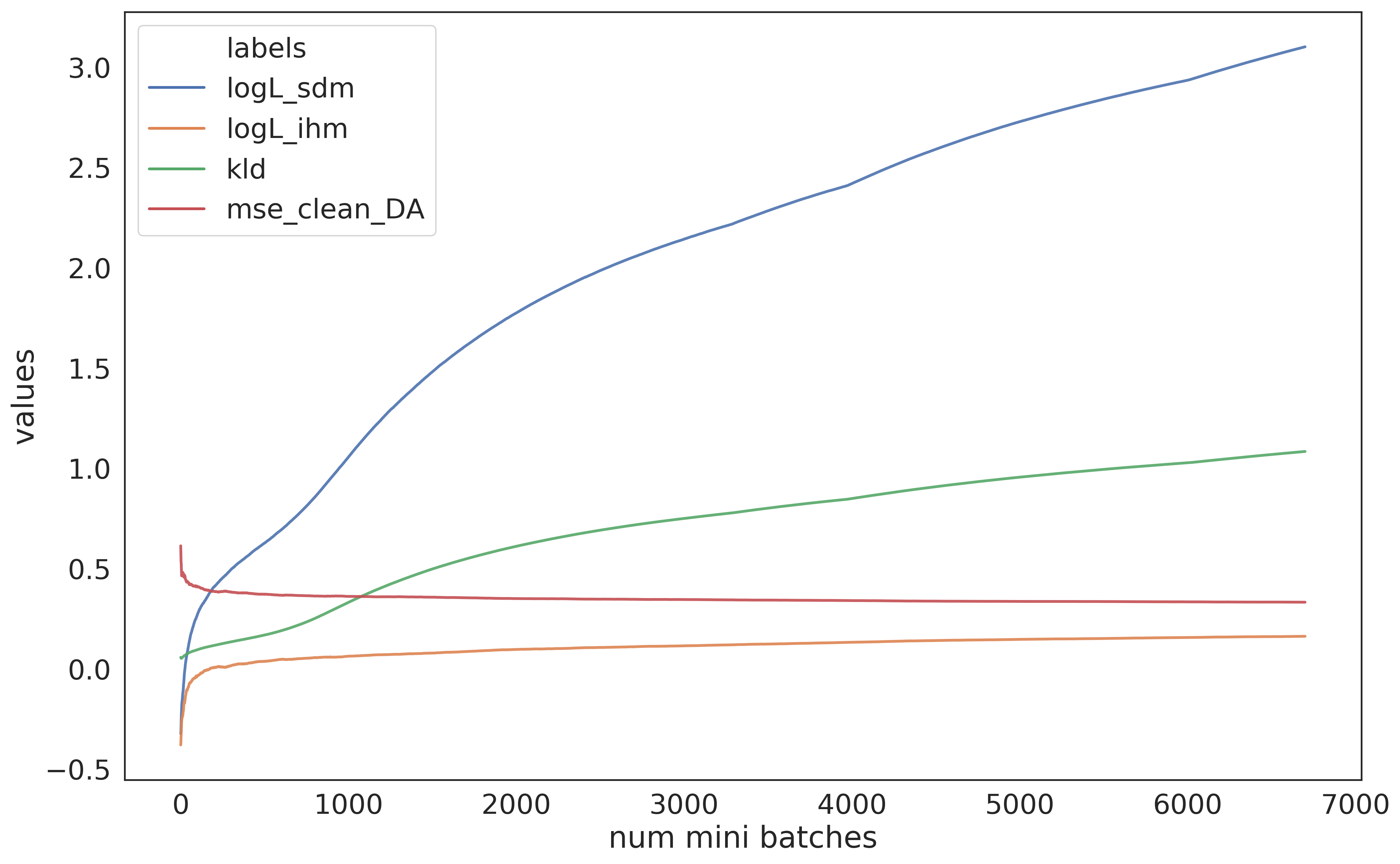}
  \caption{plot of individual losses in joint VAE vs number of mini-batches}
  \label{fig:thePlot1}
\end{figure}
\begin{figure}[t]
  \centering
  \includegraphics[width=\linewidth]{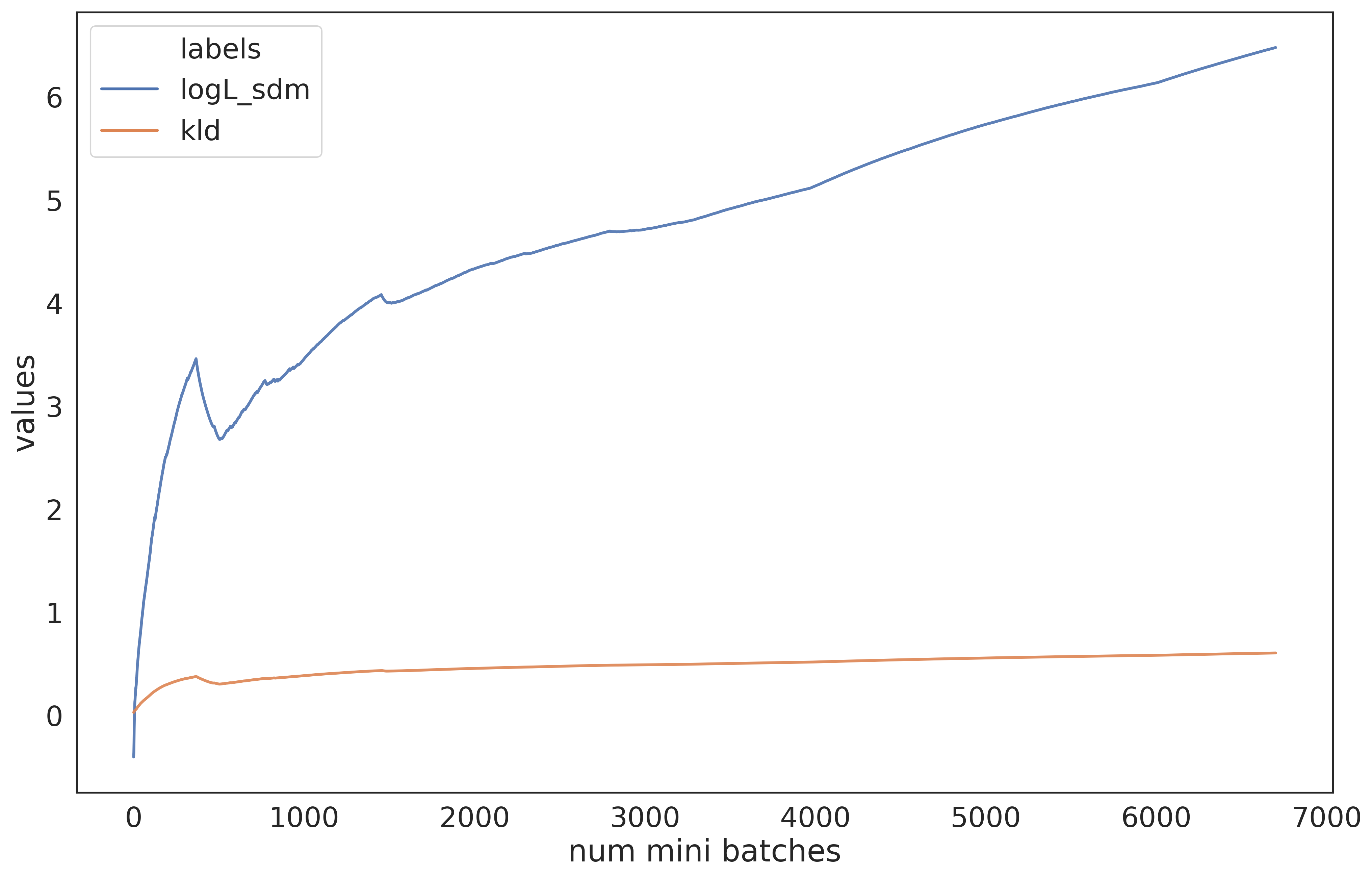}
  \caption{plot of individual losses in standard VAE vs number of mini-batches}
  \label{fig:thePlot2}
\end{figure}

In order to understand the efficacy of joint VAE, we plot individual terms in loss given by Eq~\ref{eq:loss-proposed-woApprox}, with all hyperparameters equal to $1$, in Figure~\ref{fig:thePlot1}.
Here ``logL'' refers to log-likelihood of the current batch being processed and calculated by taking negative of $\text{MSE}_\text{x}$ and $\text{MSE}_\text{y}$ in Eq~\ref{eq:loss-proposed-woApprox}.
For comparison, we plot similar terms in ELBO of standard VAE, given by Eq~\ref{eq:lowerbound-vae}, in Figure~\ref{fig:thePlot2}.
This standard VAE is trained directly on SDM features.
It can be seen from the figures that log-likelihood of SDM features is much lower in joint VAE which can be attributed to the fact that latent space in joint VAE has to learn characteristics of both SDM and IHM.
Similarly, values of KLD is higher in joint VAE to propel better reconstruction of IHM features.
As expected, MSE corresponding to DA on encoder side decreases initially then saturates.
Overall, these observations support the reported results.

\section{Conclusions}
\label{sec:conclusions}
In this paper, we explore joint VAE for far-field speech enhancement in matched scenario.
We show that when joint VAE based mapping from far-field to close-talk features is trained jointly with AM, a significant improvement is observed compared to mapping using DA.
We further extend joint VAE by relaxing the assumptions made in posterior distribution and propose a suitable architecture for the same.
Experimental results on AMI corpus show that the proposed method yields an absolute improvement of $2.5\%$ in WER compared to DA based speech enhancement model trained in matched scenario.
It is also shown that joint VAE outperforms AM trained directly on SDM features by an absolute margin of $3.96\%$ in WER.
We also deduce the best set of hyperparameters involved in the final loss of joint VAE trained with AM, empirically.
Surprisingly, prioritising reconstruction of close-talk features or cross entropy loss does not yield the best result. Instead, equal value of all hyperparameters yields the best WER.
\bibliographystyle{IEEEtran}
\bibliography{refs,strings}

\end{document}